\documentclass[reprint,amsmath,amssymb,aps,showpacs, pra]{revtex4-1}

\usepackage{graphicx}
\usepackage{dcolumn}
\usepackage{bm}

\pdfoutput=1
\usepackage[english]{babel}
\usepackage{hyperref}
\usepackage{amssymb}
\usepackage{graphicx}
\usepackage{comment}
\usepackage[titletoc,toc,title]{appendix}
\usepackage[colorinlistoftodos]{todonotes}
\usepackage[utf8]{inputenc}
\usepackage{bm}
\usepackage{siunitx}
\DeclareSIUnit\gauss{G}
\usepackage{amsmath}
\usepackage{mathrsfs}
\usepackage{braket}
\usepackage{physics}
\usepackage{leftidx}
\usepackage{wrapfig}
\usepackage[retainorgcmds]{IEEEtrantools}

\definecolor{darkCyan}{rgb}{0, 0.5, 0.5}
\definecolor{lightGrey}{rgb}{0.5, 0.5, 0.5}

\newcommand{\Rb}[1]{$^{#1}$Rb}
\newcommand{\gF}{\ensuremath{g_F}}
\newcommand{\muB}{\ensuremath{\mu_B}}

\newcommand{\refsec}[1]{Sec.~\ref{#1}} 
\newcommand{\reffig}[1]{Fig.~\ref{#1}} 
\newcommand{\rf}[0]{RF} 
\newcommand{\drf}[0]{\ensuremath{\omega}} 
\newcommand{\rabi}[0]{\ensuremath{\Omega}} 
\newcommand{\w}[1]{\ensuremath{\drf_{#1}}} 
\newcommand{\W}[1]{\ensuremath{\rabi_{#1}}} 
\newcommand{\eqnref}[1]{Eq.~\ref{#1}}

\begin{document}

\preprint{APS/123-QED}

\title{Ultracold atoms in multiple-radiofrequency dressed adiabatic potentials}
\author{T.L.~Harte$^1$, E.~Bentine$^1$, K.~Luksch$^1$, A.J.~Barker$^1$, D.~Trypogeorgos$^2$, B.~Yuen$^1$}
\author{C.J.~Foot$^1$} 
\email{christopher.foot@physics.ox.ac.uk}
\affiliation{$^{1}$ Clarendon Laboratory, University of Oxford, Oxford OX1 3PU, United Kingdom \\$^{2}$ Joint Quantum Institute, University of Maryland and National Institute of Standards and Technology, College Park, MD 20742, United States of America}

\date{\today}

\begin{abstract}
We present the first experimental demonstration of a multiple-radiofrequency dressed potential for the configurable magnetic confinement of ultracold atoms. 
We load cold \Rb{87} atoms into a double well potential with an adjustable barrier height, formed by three
radiofrequencies applied to atoms in a static quadrupole magnetic field. 
Our multiple-radiofrequency approach gives precise control over the double well characteristics, including the depth of individual wells and the height of the barrier, and enables reliable transfer of atoms 
between the available trapping geometries.  
We have characterised the multiple-radiofrequency dressed system using radiofrequency spectroscopy, finding good agreement with the eigenvalues numerically calculated using Floquet theory. 
This method creates trapping potentials that can be reconfigured by changing the amplitudes, polarizations and frequencies of the applied dressing fields, and easily extended with additional dressing frequencies.
\end{abstract}

\maketitle

\section{Introduction}
Our understanding of quantum systems has been shaped 
by the ability to study ultracold atoms in a variety of trapping geometries. 
These range from regular potentials such as lattices~\cite{Bloch2012}, waveguides~\cite{Navez2016}, rings~\cite{Ramanathan2011, Jendrzejewski2014} and box traps~\cite{Gaunt2013, Chomaz2015} to more arbitrary configurations such as tunnel junctions~\cite{Husmann2015} or disordered potentials~\cite{Choi2016}.

Such traps are often implemented using optical methods, exploiting their versatility  
in spite of drawbacks such as unwanted corrugations from fringes, sensitivity to alignment and off-resonant scattering processes that require large detunings and associated optical powers.

The application of a radiofrequency (\rf{}) field to a static magnetic trap 
dramatically changes the character of the confinement~\cite{Zobay2001, Colombe2004}, providing additional parameters to control the potential while retaining the advantages over optical dipole force traps. 
A single \rf{} applied on an atom chip \cite{Reichel2011} has been used to coherently split a 1D quantum gas~\cite{Schumm2005}, a technique since used to shed light on the nature of thermalisation in near-integrable 1D quantum systems \cite{Gring2012}. \rf \ `dressed' adiabatic potentials (APs) have also been employed to probe 2D gases \cite{Merloti2013, DeRossi2016}. Ring traps can be implemented by time averaging \cite{Lesanovsky2007,Gildemeister2010} or by adding an optical dipole potential \cite{Heathcote2008}, and are used to study superflow or for matter-wave Sagnac interferometry~\cite{Navez2016}. The introduction of a multiple-radiofrequency (MRF) field provides an additional means by which to shape these potentials~\cite{Courteille2006}, further increasing the versatility of magnetic traps.

In this work we demonstrate MRF APs for the first time, creating a highly configurable double well potential with three radiofrequencies.
Dynamic control over these potentials, which take the form of two parallel sheets, can be achieved by manipulating the \rf{} polarisation and amplitude, or properties of the underlying static field~\cite{Hofferberth2006,Gildemeister2012,Navez2016}.  
These traps are intrinsically state- and species-selective~\cite{Extavour2006,Courteille2006, Bentine2017}, with demonstrably low heating rates when created using macroscopic coils located a few cm from the atoms~\cite{Merloti2013}.
Magnetic double well potentials have previously been demonstrated using a single \rf{} on an atom chip~\cite{Schumm2005,Hofferberth2006}, and by time-averaging either a bare magnetic trap~\cite{Thomas2002,Tiecke2003} or AP~\cite{Lesanovsky2007,Gildemeister2010};
our MRF method builds upon these works to offer increased tuneability through independent control of the constituent dressing field components.
This double well potential could be developed to investigate 
tunnelling dynamics or cold-atom interferometry~\cite{Lesanovsky2006,Schumm2005} between pairs of 2D sheets. As a natural extension, additional frequency components can be applied to produce lattices~\cite{Courteille2006}, continuous potentials, or wells connected to a reservoir~\cite{Hunn2013}.

Our discussion begins with an introduction to the theory of MRF dressed potentials in~\refsec{sec:theory}, focusing on the  experimentally demonstrated three-frequency field. In \refsec{sec:Experiment} we present our experimental results, exploring the manipulation of atoms in our MRF double well potential. We describe the experimental apparatus and methods in \refsec{sec:ExpMethods} and demonstrate precise control over the potential landscape in \refsec{sec:shaping}. 

After a discussion of \rf{} spectroscopy methods in \refsec{sec:singleRFSpec}, we use this technique to probe the MRF potential landscape and validate our theoretical model in \refsec{sec:MRFspec}.
We conclude in \refsec{sec:conclusions} by outlining the new experimental possibilities arising with complex trapping geometries controlled by multiple \rf{} fields.

\section{Atoms in a multi-component \rf{} field}

The dressed-atom picture of atom-radiation interaction~\cite{Cohen-Tannoudji1969, Muskat1987} can be used to describe atoms trapped in optical, microwave~\cite{Agosta1989,Spreeuw1994}, and \rf{} fields~\cite{Zobay2001}.
An \rf{}-dressed adiabatic potential (AP) provides a trapping mechanism for cold atoms subjected to uniform \rf{} and inhomogeneous static magnetic fields~\cite{Zobay2001,GarrawayPerrin2017}. We describe the theory of MRF dressed potentials in two parts: 
\refsec{sec:AtomPhoton} presents the calculation of the quasi-energy spectrum using Floquet theory, and \refsec{sec:surfaces} describes the resulting potential surfaces and practical considerations of their implementation.

\label{sec:theory}
\subsection{Atom-photon interactions}
\label{sec:AtomPhoton}
In this work we consider $^{87}\text{Rb}$ atoms in the $F=1$ hyperfine ground state, originally confined in the static magnetic quadrupole field 
\begin{equation}
\mathbf{B}_{0}(\mathbf{r}) = B^{\prime}_{q}(x\hat{\mathbf{e}}_{x}+y\hat{\mathbf{e}}_{y}-2z\hat{\mathbf{e}}_{z})
\label{eq:quad}
\end{equation}
with $B^{\prime}_{q}$ the radial quadrupole gradient and $\hat{\mathbf{e}}_{x},\hat{\mathbf{e}}_{y},\hat{\mathbf{e}}_{z}$ the Cartesian unit vectors. 
This inhomogeneous field introduces a spatial dependence to the Zeeman splitting between hyperfine sublevels. 
We apply the homogeneous MRF dressing field
\begin{equation}
\mathbf{B}_{\text{MRF}}(t) = \sum_{j} B_{j} \frac{\cos(\omega_{j}t+\phi_{j})\hat{\mathbf{e}}_{x}-\kappa_{j}\sin(\omega_{j}t+\phi_{j})\hat{\mathbf{e}}_{y}}{\sqrt{1+\kappa_j}}
\label{eq:Bmrf}
\end{equation}
where $B_{j}$, $\omega_j$, and $\phi_j$ are the amplitude, angular frequency and relative phase of each frequency component respectively. 
In our experimental implementation we use three 
\rf{} components $\w{j} = \w{1,2,3} = 2\pi\times(5,6,7)\times\SI{0.6}{\mega\Hz}$, 
producing circularly polarised dressing fields for 
$\kappa_j=1$, and linearly polarised fields for $\kappa_j=0$.  
The following discussion describes either linear or circularly polarised \rf{} fields, for which the dressed-atom Hamiltonian of the system reads
\begin{equation} 
	V= \sum_j \hbar \omega_j a_j^{\dagger} a_j 
		+ \gF\muB \mathbf F \cdot \left[\mathbf B_0(\mathbf r) + \mathbf{B}_{\mathrm{MRF}} \right] \label{eq:VHamiltonian}
\end{equation}
where
\begin{equation}
\mathbf {F \cdot B_{\mathrm{MRF}}} = \sum_j \mathscr{E}_j \left( \frac{\alpha_j}{\sqrt{2}} F_+ + \frac{\beta_j}{\sqrt{2}} F_- + \zeta_j F_z\right) a^{\dagger} + \text{HC.}
\label{eq:VInteraction} 
\end{equation}
In this expression, 
$\mathbf{B}_{\mathrm{MRF}}(t)$ 
now describes the second quantised operator for the MRF field with mode densities $\mathscr E_j$, and amplitudes $\alpha_j,\ \beta_j$ and $\zeta_j$ as defined in Eqs.~\ref{eq:zeta} and~\ref{eq:alphabeta}. The Hermitian conjugate is indicated by HC, while \gF{} denotes the Land\'e g-factor and $\mu_B$ the Bohr magneton. 

The first term in \eqnref{eq:VHamiltonian} accounts for the energy of the \rf{} field component $j$ with angular frequency $\omega_j$ and corresponding photon creation and annihilation operators $a_j^{\dagger}$ and $a_j$.
The second term describes the interaction between the atomic spin 
$\mathbf{F}$, defined following the convention in Ref.~\cite{Foot2005}, and the total magnetic field comprising static and \rf{} components with operators 
$\mathbf B_0(\mathbf r)$ and 
$\mathbf{B}_{\mathrm{MRF}}(t)$ 
respectively.

\begin{figure*}[!t]
\centering
	\includegraphics[width=0.95\textwidth]{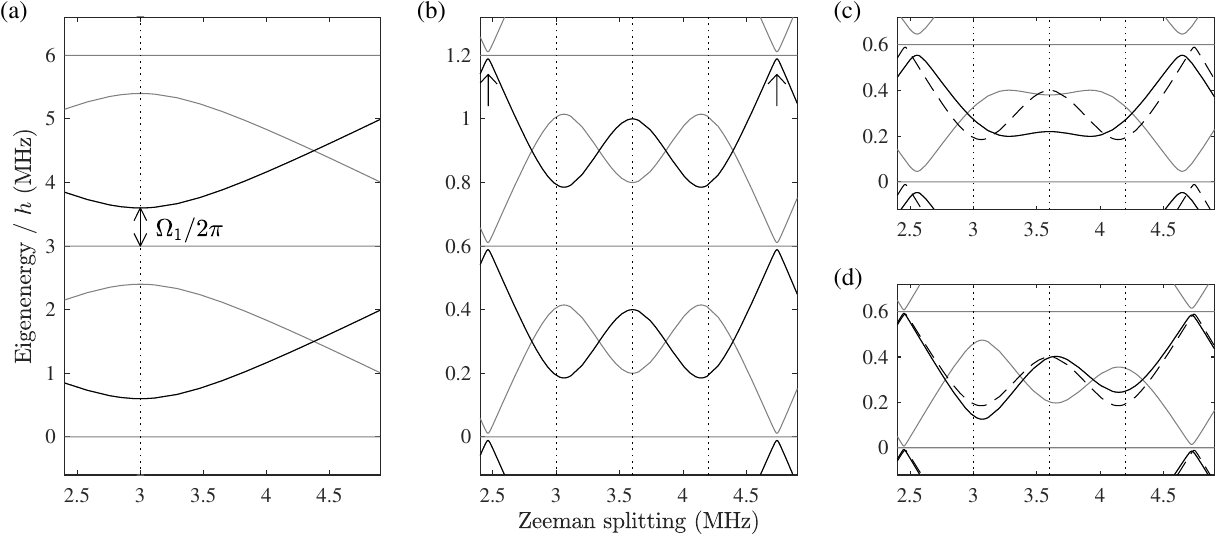}
\caption[]{
(a) Eigenenergies of the dressed atomic states showing the avoided crossing that forms due to a single applied \rf{} at \w{1} (vertical dotted line). 
The black lines show the trapping well  
for atoms in $\ket{\widetilde{m}_{F}=1}$ with Rabi frequency \W{1}. Grey lines show the untrapped $\ket{\widetilde{m}_{F}=0,-1}$ eigenenergies. 
(b) A three-component \rf{} field forms a double well. 
The system periodicity is now defined by \w{f}, the highest common factor of the \rf{} components, with 
avoided crossings formed at the resonance of each \rf{} component. 
The energies of the eigenstates at each resonance are shifted by the presence of the other \rf{} field components, translating the well minima in space.
Weak avoided crossings are also formed by multi-photon couplings at integer values of $\hbar\w{f}$, indicated by arrows. 
(c) The potential can be deformed into a broad single well by 
increasing the amplitude of the middle field component. 
Solid and dashed lines show the potential for different barrier heights. 
(d) The relative amplitudes of the outermost \rf{}s control the imbalance in well depth. Any combination of \rf{} amplitudes can be used to shape the potential, for example to mediate atom transport between the wells. Note: for clarity, we omit effects of gravity in figures except where it serves an illustrative purpose. 
}
\label{fig:theoryPlot}
\end{figure*}

The combined system of magnetically-confined atom, \rf{} radiation, and the interaction between them can be intuitively described in the dressed-atom picture, as illustrated in \reffig{fig:theoryPlot} for a single- and triple-frequency field.
In the absence of interactions with the \rf{} field, the dressed eigenstates $\ket{n_1, n_2, ..., m_F}$ are the  
tensor products of the Fock states of each \rf{} field $\ket{n_j}$ and the atomic Zeeman substates $\ket{m_F}$. 
These form a ladder of eigenenergies $\gF\muB m_F \abs{\mathbf B_0}+ \sum_j n_j \hbar \omega_j$ in which the three Zeeman substates are repeated  
with a spacing of \w{f}, the highest common factor of \rf{} photon frequencies \w{j}. The interaction described by \eqnref{eq:VInteraction} drives transitions between dressed states, turning energy level crossings into avoided crossings.

While the dressed-atom picture provides an intuitive visualisation of the \rf{} dressing process, 
the large mean photon number of the \rf{} field allows it to be represented classically 
by replacing $\mathscr E_j a_j^{\dagger}$ and $\mathscr E_j a_j$ by their mean field value $\frac12 B_j$. This is performed within the context of the interaction picture, in which $V\rightarrow U_{\mathrm {\rf{}}}^\dagger V U_{\mathrm {\rf{}}}$ and $\ket{\psi}\rightarrow U_{\mathrm {\rf{}}}^{\dagger} \ket{\psi}$ with $U_{\mathrm{\rf{}}} = \exp[i \sum_j a_j^{\dagger} a_j \omega_j t]$. 

The \rf{} field is decomposed into components parallel and perpendicular to a local axial vector $\hat{\mathbf z}'$ where $\mathbf {F \cdot B}_0=B _0 F_{z'}$. 
The parallel component is given by $\frac12 B_j \zeta_j \exp [i (\omega_j t + \phi_j)] + \text{CC}$, where CC indicates the complex conjugate, with
\begin{equation} 
\label{eq:zeta}
\zeta_j(\mathbf r) = \frac{1}{\sqrt{1+\kappa_j}} (\sin \theta \cos \phi + i \kappa_j \sin \phi). 
\end{equation}
From the definition of the static quadrupole field, 
$\cos \theta = -2 z (x^2+4 z^2)^{-1/2}$ and $\cos \phi = [(x^2+4 z^2)/(x^2+y^2+4 z^2) ]^{1/2}$. 
The anticlockwise and clockwise rotating components of the perpendicular field are $\frac12 B_j \alpha_j \exp [i (\omega_j t + \phi_j)] + \mathrm{CC}$ and $\frac12 B_j \beta_j \exp [i (\omega_j t + \phi_j)] + \mathrm{CC}$ respectively, with
\begin{IEEEeqnarray*}{rCl}
	\alpha_j (\mathbf r) &=& \frac{1}{\sqrt{2+2\kappa_j}}(\cos \theta - i \sin \theta \sin \phi - \kappa_j \cos \phi),\\
    \beta_j (\mathbf r) &=& \frac{1}{\sqrt{2+2\kappa_j}}(\cos \theta + i \sin \theta \sin \phi + \kappa_j \cos \phi). \IEEEyesnumber \label{eq:alphabeta}
\end{IEEEeqnarray*}
In this basis the semiclassical version of the Hamiltonian presented as \eqnref{eq:VHamiltonian} becomes
\begin{IEEEeqnarray*}{rl}
	V(t) =& \gF\muB B_0 F_z  \\
    &+ \frac{\gF\muB}{2} \sum_j \left[ \left( \frac{\alpha_j}{\sqrt{2}} F_- + \frac{\beta_j}{\sqrt{2}} F_+ + \zeta_j F_z \right) B_j e^{i (\omega_j t+\phi_j)} \right. \\
	&\left.	
	 +\left( \frac{\alpha_j^*}{\sqrt{2}} F_+ + \frac{\beta_j^*}{\sqrt{2}} F_- + \zeta_j^* F_z \right) B_j e^{-i (\omega_j t+\phi_j)} \right], \IEEEyesnumber \label{eq:semiclassicalH}
\end{IEEEeqnarray*}
which is periodic in time with period $T=2 \pi / \omega_f$. The coefficients $\alpha_j, \ \beta_j$ and $\zeta_j$ give the projection of the field operator in the local circular basis, with $\abs{\alpha_j}^2+\abs{\beta_j}^2+\abs{\zeta_j}^2=1$.

Using Floquet's theorem, 
the eigenstates of this time-periodic Hamiltonian, with period $T$, can be expressed in the form $\ket{\psi(t)} = \exp[i E' t/ \hbar]\ket{\Psi(t)}$, a product of a phase term and the time-periodic state vector $\ket{\Psi(t)}$, where $\ket{\Psi(0)} = \ket{\Psi(T)}$. 
Alternatively, one can write $\ket{\psi(t)} = U(t)\ket{\psi(0)}$, where $U(t)$ is the time evolution operator.
We calculate $U$ through numerical integration of the Schr\"odinger equation with the interaction Hamiltonian of \eqnref{eq:semiclassicalH}. By comparing these two equations for $\ket{\psi(t)}$, we find 
$U(T) \ket{\psi(0)} = \ket{\psi(T)} = \exp[i E' T/\hbar] \ket{\psi(0)}$. The phases $E'T/\hbar$ can be associated with the energy of the dressed eigenstates of \eqnref{eq:VHamiltonian} at time $T$~\cite{Shirley1965, Yuen2017} 
such that the dressed state eigenenergies modulo $\hbar\omega_f$ are given by the $2F+1$ 
eigenvalues of $(-i\hbar/T) \log U(T)$. These eigenenergies are illustrated in~\reffig{fig:theoryPlot} for the three-\rf{} example that we investigate experimentally. 

\subsection{Adiabatic potentials}
\label{sec:surfaces}
The interaction $\gF\muB \mathbf {F \cdot B}_{\mathrm {MRF}}$ couples the states to form avoided crossings 
at values of the static field for which the energy splitting $\gF\muB B_{0}$ is resonant with an integer multiple of $\hbar\w{f}$. 
When this interaction is sufficiently strong and the static field orientation varies sufficiently slowly with position, an atom traversing an avoided crossing can adiabatically follow this new eigenstate, labelled by the quantum number $\widetilde{m}_{F}$~\cite{Lesanovsky2006a}.

In the case of a single applied \rf{} with angular frequency \w{1} shown in \reffig{fig:theoryPlot} (a), atoms trapped in 
$\widetilde{m}_{F}=1$ 
experience a trapping potential 
$U_{AP}(\mathbf{r})=\widetilde{m}_{F}\hbar\sqrt{\delta^{2}(\mathbf{r})+\Omega^{2}_{1}(\mathbf{r})}$ 
where 
$\delta(\mathbf{r}) = |\gF\muB B_{0}(\mathbf{r})/\hbar|-\w{1}$ 
gives the angular frequency detuning of the \rf{} from resonance and the Rabi frequency is determined by the applied \rf{} amplitude and polarisation.

The spatial variation of the static field amplitude $B_{0}(r)$ translates the detuning-dependence of the potential to a spatial dependence, such that for the static quadrupole of~\eqnref{eq:quad} the resultant trapping potential forms an oblate spheroidal `shell trap'. 
Atoms are trapped on the surface of this resonant spheroid, over which the spatial variation of the coupling strength is dictated by the \rf{} polarisation. 

The Rabi frequency for a circularly polarised \rf{} field is given by
\begin{equation}
\W{1} = \frac{\gF\muB B_{1}}{2 \sqrt{2} \hbar}\left(1 \pm \frac{2z}{\sqrt{x^2+y^2+4z^2}} \right)
\label{eq:RabiCirc}
\end{equation}
with $B_{1}$ the magnetic field amplitude of the \w{1} \rf{} field and $x,y,z$ Cartesian coordinates with an origin at the centre of the quadrupole field. 
The sign of the second term depends on the handedness of the \rf{} field polarisation; 
in this work the handedness is chosen such that the coupling is maximised at the south pole of the resonant spheroid. 
For the case of an \rf{} field linearly polarised in the $xy$ plane the Rabi frequency instead takes the form
\begin{equation}
\W{1} = \frac{\gF\muB B_{1}}{2\hbar}
\left( \frac{r_{\perp}^{2} + 4z^{2}}{r_{\perp}^{2}
+r_{\parallel}^{2} + 4z^{2}} \right)^{1/2}
\label{eq:RabiLin}
\end{equation}
where $r_{\parallel}$ and $r_{\perp}$ describe the coordinates parallel and perpendicular to the polarisation direction of the linear \rf{} field. 
The resonant spheroid therefore has maximum coupling at points for which the parallel component is zero, and zero coupling at the points on the equator for which the perpendicular component is zero. 

As illustrated in~\reffig{fig:theoryPlot} (b), this principle can be easily extended to the MRF case, 
in which the three first-order avoided crossings form two trapping wells separated by an anti-trapping barrier for an atom in $\widetilde{m}_{F}=1$. 
This results in trapping on two concentric spheroids forming a spatially-extended double well 
in which the relative heights of the barrier and both wells are controlled by the three separate input \rf{}s. 
Multi-photon interactions lead to cross-talk between these features, and the impact of the amplitude \W{j} of each avoided crossing on the properties of its neighbours is investigated experimentally in \refsec{sec:shaping} and~\ref{sec:MRFspec}. Also studied in \refsec{sec:MRFspec} is the effect of the relative phase $\phi_{j}$ between \rf{} components; this alters the overall shape of the MRF waveform and thus influences the strength of nonlinear multi-photon processes that occur.

Adiabaticity constraints motivate 
the choice of parameters including the frequency separation, \rf{} amplitudes and static field gradient.   
An atom with constant velocity $v$ moving through this spatially-varying potential will remain trapped 
with a probability approximately given by the Landau-Zener model: this states that 
$P_{\text{LZ}} = ( 1-\exp[- h \Omega^{2}/ 4 \gF\muB\partial_{t}B_{0}(vt)] )^2$
where the time derivative of the static field $B_0$ indicates the field gradient as experienced by the moving atom~\cite{Courteille2006,Burrows2017}. 
Minimising the well spacing requires a dressing \rf{} frequency separation comparable to the Rabi frequency of each \rf{} component. 

As the piecewise approach presented in Ref.~\cite{Courteille2006} is invalid in this limit~\cite{Morgan2014} \cite{Chakraborty2017ARf-fields}, Floquet theory is employed to calculate the MRF dressed state eigenenergies. 
Numerical artefacts are removed by appropriate meshing over the range of magnetic field values considered, while an intuitive depiction of MRF dressing that uses the resolvent formalism to discard these artefacts is explored in Ref.~\cite{Yuen2017}.

\section{Experimental implementation of the MRF potentials}
\label{sec:Experiment}
\subsection{Trapping atoms in an adiabatic potential}
\label{sec:ExpMethods}
In standard operation, we routinely produce BECs of $3.5\times10^{5}$ \Rb{87} atoms in the $\ket{F=1, m_{F}=-1}$ hyperfine state using a time-orbiting potential (TOP) trap~\cite{Petrich1995}, via an experimental sequence that we can truncate to load thermal atoms into an AP prior to a final stage of evaporation. The TOP is formed by applying a bias field, rotating at \SI{7}{\kilo\Hz}, to the static quadrupole field of~\eqnref{eq:quad}. 
This bias field sweeps the quadrupole field in a horizontal circular orbit with a rotation radius given by $B_{T}/B_{q}^{'}$, with $B_{T}$ the amplitude of the TOP field.

\begin{figure}[!t]
\includegraphics[width=0.4\textwidth]{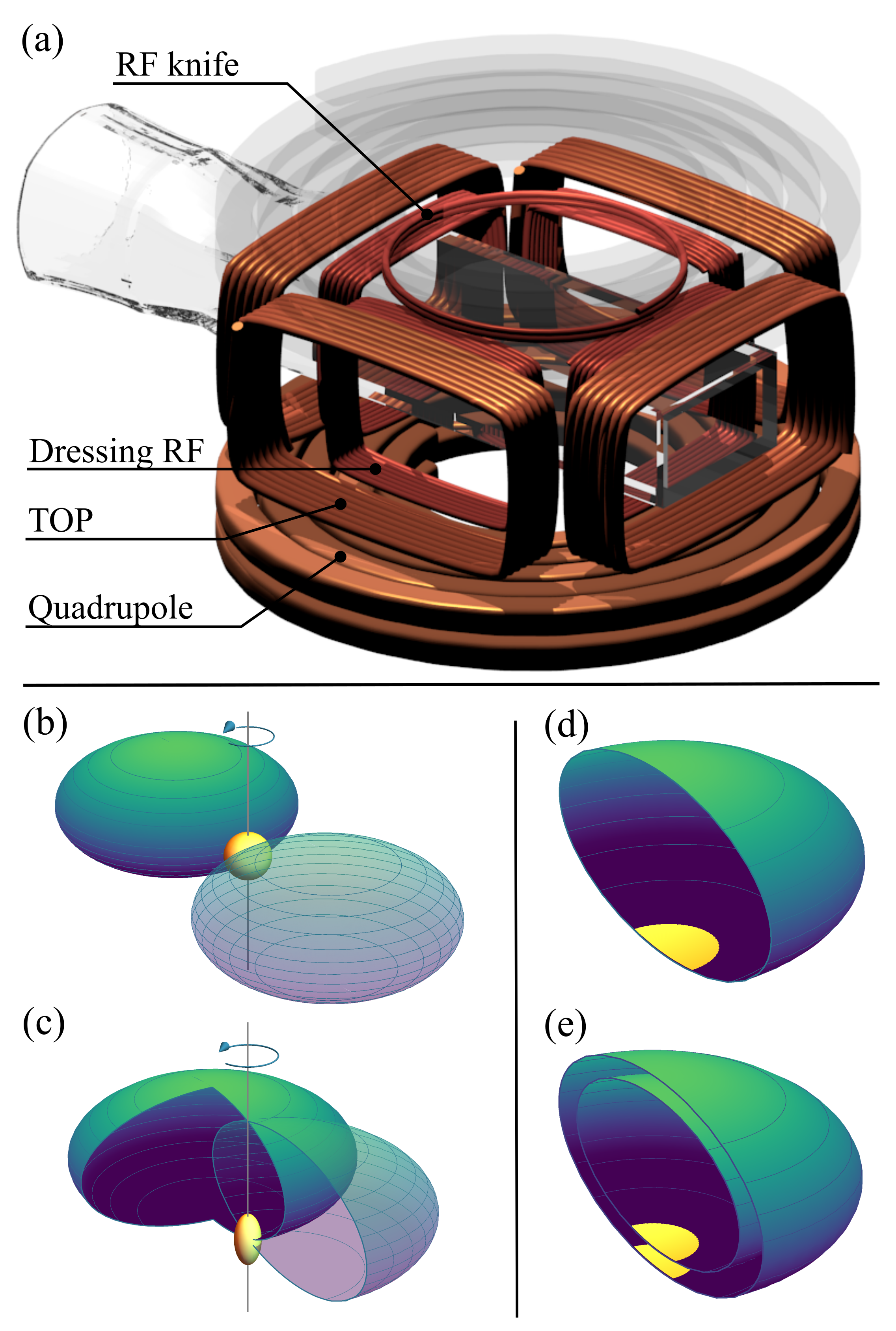}
\caption{(a)
Our magnetic fields are generated by a coil array that surrounds the ultra-high vacuum glass cell. 
The racetrack coils are connected in pairs to generate orthogonal components of the rotating TOP and dressing \rf{} fields. The single circular coil provides the \rf{} knife used in evaporation and spectroscopy. 
Quadrupole gradients are applied using the large anti-Helmholtz coils located above and below the AC array. 
(b)
Prior to loading the AP, the atoms are confined in the TOP. When the dressing \rf{} is applied, the rotating bias field of amplitude $B_{T}$ moves the resonant spheroid in a circular orbit. 
For $B_{T} > \hbar\w{1} / \mu_B g_F$ the shell orbits outside the atom cloud (yellow). 
(c) Lowering $B_{T}$ causes an intersection of spheroid and rotation axis, creating and loading the TAAP where \rf{} evaporation can be performed. 
(d) $B_{T}\to 0$, loading atoms into the shell trap. 
(e) Applying three dressing \rf{}s creates the double-shell potential.
}
\label{fig:setup}
\end{figure}

The TOP and dressing \rf{} fields are generated by a coil array that surrounds the atoms, with an extent of a few cm. 
This array is illustrated in~\reffig{fig:setup}.
The \rf{} signals for each coil and frequency component are independently generated by direct digital synthesis (DDS)~\footnote{Analog Devices AD9854}. 
This digital control over the amplitude and polarisation of each dressing field component enables us to precisely sculpt the waveform and resultant potential as a function of time. 
The signals for each coil are combined using splitters~\footnote{Mini-Circuits ZSC-2-2}, 
and amplified by 25~W amplifiers~\footnote{Mini-Circuits LZY-22+}. 
The \rf{} coil array has a self-resonance of approximately \SI{7}{\mega\Hz} such that,  with a custom wideband impedance match, we can confine atoms in APs with dressing frequencies in the range $2\pi\times2.7$ to $2\pi\times4.4$~MHz without additional amplification.
Mixing processes in the amplifiers constrain us to use only combinations of dressing frequencies with a common fundamental $\w{f}$,  
ensuring that the resulting intermodulation products are far detuned from transitions between dressed states such that we avoid losses. 

We load a single-\rf{} shell with thermal atoms as described  in~\cite{Gildemeister2010,Sherlock2011,Gildemeister2012}, combining the 
TOP field with dressing \rf{} to produce a time-averaged adiabatic potential (TAAP) as illustrated in~\reffig{fig:setup}(b-e). 
The dressing \rf{} is switched on while the TOP field satisfies $2\hbar\w{1}/g_{F}\mu_{B}>B_{T}>\hbar\w{1}/g_{F}\mu_{B}$ such that 
the TOP field sweeps the resonant spheroid in an orbit outside the location of the atom cloud. 
With an \rf{} amplitude on the order of $\W{j}=g_{F} \mu_{B} B_{j}/(\sqrt{2} \hbar) = 2\pi\times400$~kHz at the south pole of the spheroid, 
decreasing $B_{T}$ allows us to load the atoms into the TAAP formed at the 
lower of the two intersections of the spheroid with the rotation axis under the influence of gravity. 
The \rf{} field is circularly polarised in the laboratory frame, with a handedness that maximises the interaction strength at the bottom of the resonant spheroid.
Using an additional weak field we then optionally perform forced \rf{} evaporation to BEC in \SI{2}{\second}, exploiting the enhanced radial trap frequencies inherent to the TAAP.
Reducing $B_{T}$ to zero subsequently loads atoms from the TAAP onto the lower surface of the shell. 
This reliably loads condensates of greater than 
$3\times 10^{5}$ 
atoms into the shell trap with negligible heating. 

\subsection{Potential shaping and the double shell}
\label{sec:shaping}
\begin{figure*}
\includegraphics[width=0.88\textwidth]{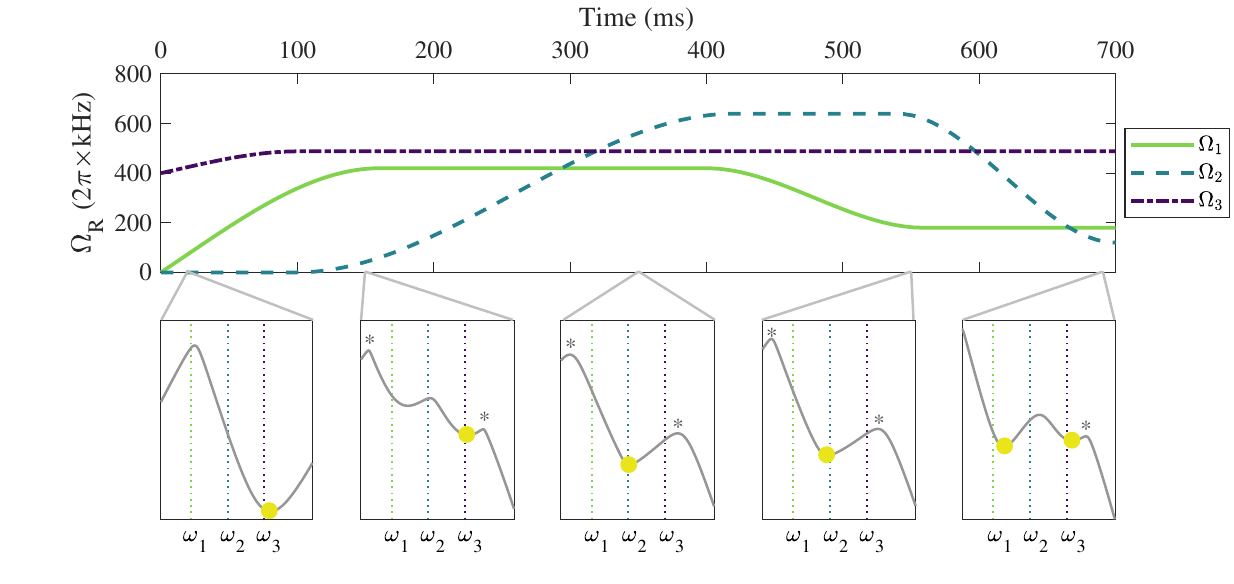}
\caption{A typical time sequence of the dressing \rf{} amplitudes used to load a BEC into the double shell configuration, where $\W{j}=g_{f}\mu_{B}B_{j}/(\sqrt{2}\hbar)$ 
denote the constituent field amplitudes. 
The lower panels show the three-frequency potential (including gravity) at key times during this loading sequence, with dotted lines indicating the locations associated with the first-order resonances of the dressing frequencies, \w{1}, \w{2} and \w{3}. 
This illustrates the transformation into a three-frequency single well before the barrier is raised to split the cloud between the two wells. 
These experimental parameters avoid losses due to the 
second-order resonances indicated by asterisks in the panels above.  
The relative amplitudes of the \rf{} components determine the final distribution of atoms between each well. 
}
\label{fig:LoadMRF}
\end{figure*}

This single-\rf{} configuration forms the starting point for the MRF double well potential, with atoms initially confined in the shell corresponding to either \w{1} or \w{3} and ultimately transferred into the combined 
$\w{1,2,3} = 2\pi\times(5,6,7)\times\SI{0.6}{\mega\Hz}$ potential.
In our apparatus the $2\pi\times0.6$ MHz frequency difference between \rf{} components maps to a spatial well separation of $\sim140$\si{\micro\m}
at a quadrupole gradient $B_{q}^{'} = 62.45$ G/cm,
allowing the trapping wells to be clearly resolved with our low-resolution imaging system.
The double shell loading procedure is shown in \reffig{fig:LoadMRF} for the case of loading from a single shell at \w{3}. 
We first ramp up \W{1}, 
which has a minimally perturbative effect on the potential near the atoms but establishes this resonance in preparation for the subsequent 
application of the field at \w{2}. 
As 
shown in \reffig{fig:theoryPlot}, 
the avoided crossing formed by \w{2} takes the form of an anti-trapping barrier. As $\rabi_2$ increases, the barrier is lowered and 
the MRF potential is flattened, rounded out, or tilted slightly according to the desired loading scheme and relative values of \W{1}, \W{2} and \W{3}. 
To minimise any sudden changes in the width of the potential experienced by the atoms as the barrier is lowered, \W{1} is held at an artificially high value, and lowered to the value at which atoms can be transferred only once the barrier has been ramped down fully. 
Once atoms equilibrate within this new potential, we raise the barrier to separate the wells and complete the loading process. This method is illustrated in~\reffig{fig:LoadMRF} for the \rf{} ramps used to split a BEC between the two shells, and 
variants on this loading scheme were used in the remaining figures. 
The second-order resonances apparent in \reffig{fig:LoadMRF} place an upper limit to the well depth of $\hbar \omega_f$;   
the combination of \rf{} amplitudes and frequency separation 
are therefore chosen to complement the temperature of atoms loaded into the potential. 

\begin{figure}
\centering
\includegraphics[width=0.47\textwidth]{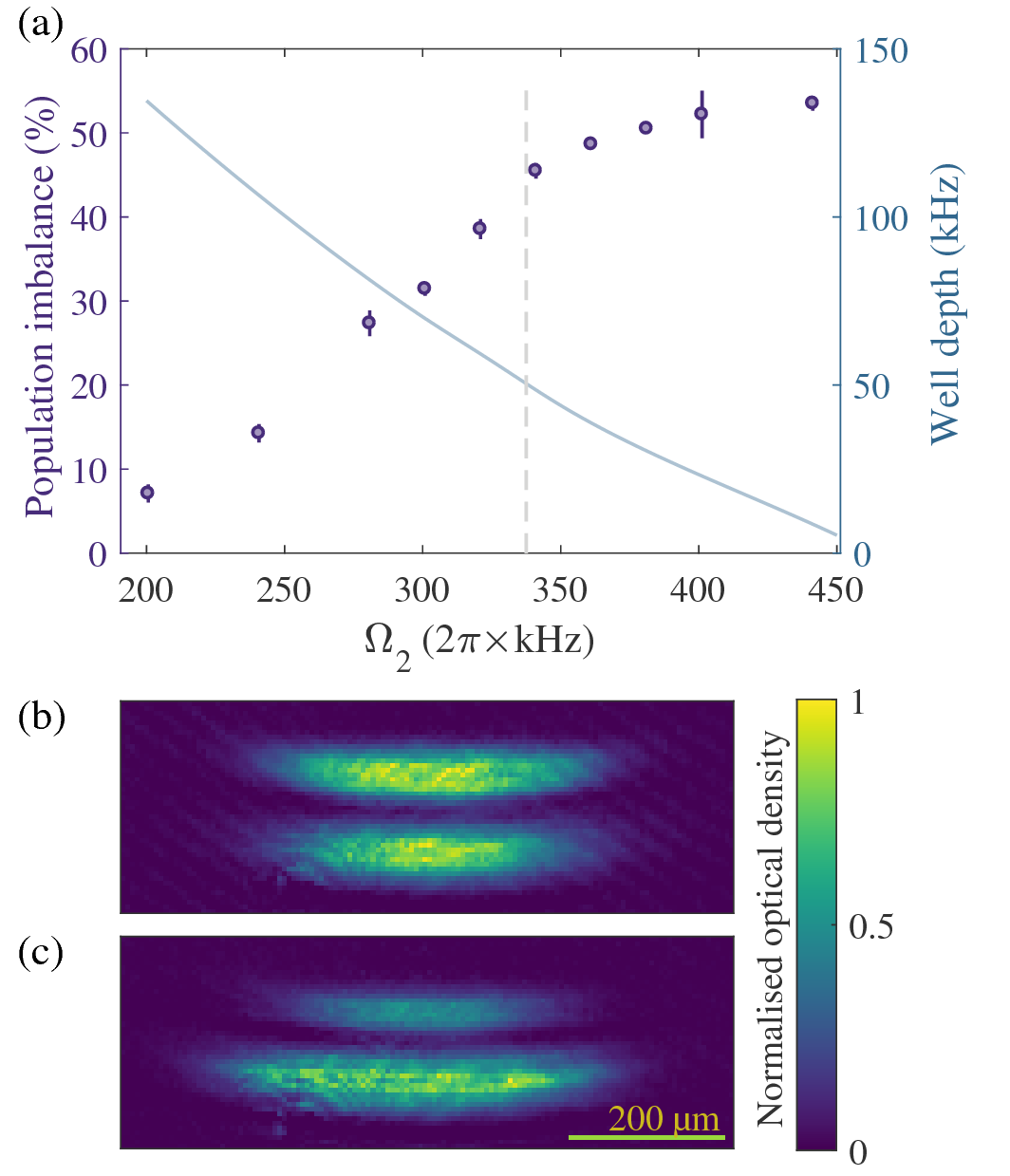}
\caption{(a) The percentage of atoms loaded 
from the wells formed by \w{3} to \w{1} 
for a given maximum amplitude of the \w{2} field, expressed in terms of $\W{2}=g_{F}\mu_{B}B_{2}/(\sqrt{2}\hbar)$ (purple dots). The \rf{} amplitude ramps are qualitatively similar to \reffig{fig:LoadMRF}, with 
$\W{1} = 2\pi\times \SI{192}{\kilo\Hz}$ and $\W{3} = 2\pi\times\SI{442}{\kilo\Hz}$. 
This amplitude disparity compensates the effects of gravity, with a quadrupole gradient of $B_{q}^{'} =$ \SI{154}{\gauss\per\centi\meter}.
The barrier was ramped to its maximum value over \SI{400}{\milli\second}, then reduced to 
$2\pi\times\SI{90}{\kilo\Hz}$ over \SI{100}{\milli\second}. The blue line shows the effective well depth (right hand scale) seen by atoms in the well at \w{3} for each final value of \W{2}, and the dashed vertical line indicates the barrier height for which a separate well at \w{1} can no longer be resolved. 
(b), (c) Absorption images of thermal atoms in the double shell at a quadrupole gradient of \SI{60}{\gauss\per\centi\meter} after 1 ms time of flight, with (b) an approximately balanced configuration with \SI{52}{\percent} of atoms in the upper shell and (c) \SI{75}{\percent} of the population in the lower shell. 
The colour bar indicates the colour map used for all absorption images in this work, and has a linear scaling from 0 to the maximum optical depth in each image. 
}
\label{fig:BarrierLoad}
\end{figure}

The final population imbalance between the wells is influenced by 
the relative amplitudes of each \rf{} component during the ramp. 
The effect of barrier height is illustrated in \reffig{fig:BarrierLoad}, where we vary the maximum value of \W{2} to load a controllable proportion of atoms between the lower and upper wells, formed by \w{3} and \w{1} respectively. 
Starting from a cloud of thermal atoms in the lowest shell, the \rf{} components \w{1} and \w{2} are turned on adiabatically following a similar procedure to that described in~\reffig{fig:LoadMRF} in which \W{1} is ramped directly to its final value. 
Initially, few atoms possess sufficient energy to cross the high barrier that results from a small \W{2}, and minimal population redistribution between the wells occurs. 
Increasing $\rabi_2$ to lower the barrier allows more atoms to populate the second well. At around $\rabi_2 = 2\pi\times400$ kHz the barrier vanishes and the atoms distribute themselves across the broad single well formed by the three \rf{} dressing frequencies as shown in \reffig{fig:theoryPlot}(c).
Finally, $\rabi_2$ is decreased to raise the barrier and split the population distribution into two distinct wells, with the proportion reflecting any imbalance between the lowest energy of each well. 
Figure~\ref{fig:BarrierLoad}(a) illustrates a loading process that transfers 52\% of the atoms into the well defined by \w{1}. 
This could be corrected or exacerbated by adjusting either $\rabi_1$ or $\rabi_3$ to raise or lower the potential energy minimum of each well.

\begin{figure}[!t]
\centering
\includegraphics[width=0.5\textwidth]{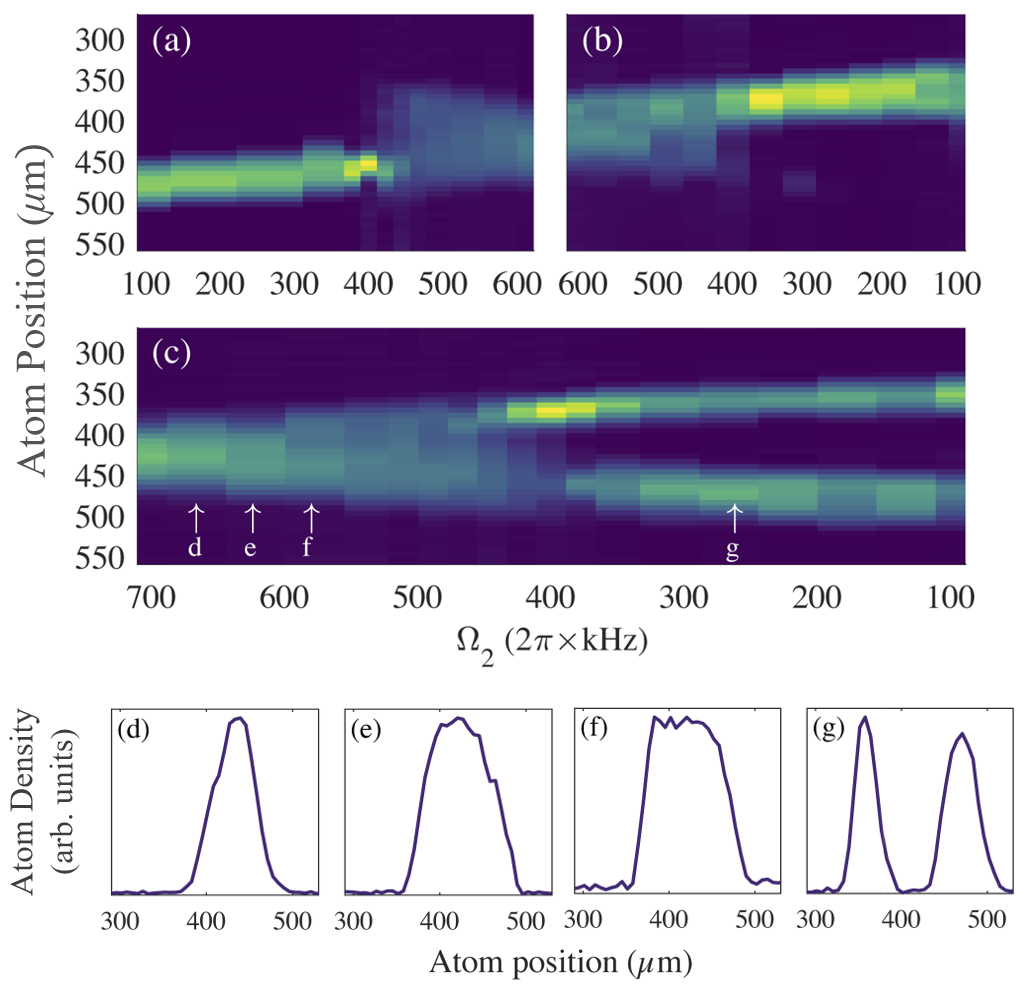}
\caption{(a)-(c) Vertical slices through in-trap absorption images of the MRF potential plotted against barrier amplitude $\rabi_2$ and averaged over several experimental runs.
Displacement is measured from the centre of the quadrupole trap, and each slice scaled to the same total atom number.
(a) All atoms begin in the shell at \w{3}. With $\Omega_{1} = \SI{192}{\kilo\Hz}$ and $\Omega_{3} = \SI{446}{\kilo\Hz}$, \W{2} is ramped up (lowering the barrier) to flatten the potential and load the atoms into a broad single well. 
(b) A transport sequence with $\Omega_{1} = 2\pi\times192$~kHz and $\Omega_{3} = 2\pi\times511$~kHz. 
This tips atoms across the flattened 3-\rf{} potential to load atoms from the lower to the upper shell upon reducing \W{2} to raise the barrier.  
(c) Loading a double-shell configuration from the flattened 3-\rf{} potential, with $\Omega_{1} = 2\pi\times192$~kHz and $\Omega_{3} = 2\pi\times446$~kHz to maintain approximately equal atom populations in each well. The highest values of \W{2} correspond to a single well, with two distinct wells forming as \W{2} is reduced to raise the barrier.
The apparent transfer of atoms into the shell at \w{1} around $\Omega_{2}=2\pi\times400$~kHz is a normalisation artefact, resulting from atom loss from the lower well due to technical noise in the apparatus.
(d)-(g) show the line plots of atom density for the snapshots in the double shell load sequence at barrier amplitudes marked in (c) and corresponding to $\W{2}=2\pi\times660, 622, 577, 266$~kHz for (d)-(g) respectively. This shows the progression from 3\rf{} single well (d) and (e) to flat-bottomed `box trap' (f) and double-shell potential (g). 
}
\label{fig:wellPositions}
\end{figure}

Figure~\ref{fig:wellPositions} illustrates the atom density arising from two possible transport sequences. 
Keeping the lowest energies of each well approximately equal allows us to load the balanced double shell with approximately \SI{75}{\percent} efficiency in atom number, 
while deliberately mismatching these energies allows a full population transfer between the wells. 
Crucially,~\reffig{fig:wellPositions} also demonstrates the effect of the barrier amplitude on the positions of the two trapping wells that is shown in the calculated energy levels in \reffig{fig:theoryPlot}: the \w{1} and \w{3} potential minima are drawn closer together as the barrier is lowered to form the broad single well.

The simple potential shaping schemes demonstrated here for three frequencies comprise single wells, a double well, and a flattened three-frequency well. 
We have also demonstrated a method of dynamic control that provides the intermediate stages for loading. This approach can be extended 
in a straightforward manner by applying additional dressing \rf{}s.

\subsection{RF spectroscopy} 
\label{sec:singleRFSpec}
RF spectroscopy is an experimental technique commonly used to precisely characterise bare magnetic traps and adiabatic potentials~\cite{Hofferberth2007, Easwaran2010}. 
A weak probe \rf{} is applied to atoms held within the trap, causing expulsion of atoms 
when the probe \rf{} is resonant with a transition between trapped and untrapped states. 
With these resonances appearing as dips in the measured atom number,
the probe frequency is varied to map out the spectrum of transitions.
For a BEC, this resonance has a width on the order of the chemical potential (typically \si{\kilo\Hz}) 
while for a thermal cloud the resonance is broadened due to the thermal distribution of atoms in the trap\,\cite{Easwaran2010}.

RF spectroscopy is employed here to characterise the key components of our trapping fields: the TOP field magnitude $B_{T}$, 
amplitudes of applied dressing \rf{} components, and ultimately the MRF eigenenergies. 
$B_{T}$ is measured by \rf{} spectroscopy of a condensate confined in the TOP, and $B_{q}^{\prime}$ calibrated by measuring the trap frequency of the centre of mass mode of a condensate oscillating in this approximately harmonic potential for a known current through the quadrupole coils.

To calibrate the \rf{} amplitudes, 
transition frequencies are measured for 
single-\rf{} shells at $\drf_{1,2,3}=2\pi\times (3, 3.6, 4.2)$ \si{\mega\Hz}. 
We use linearly polarised \rf{} to measure the \rf{} fields in x and y directions independently. 
The Rabi frequencies are calculated from these measured resonances through Floquet theory as described in \refsec{sec:theory}. 
This calculation incorporates the Bloch-Siegert shifts~\cite{Bloch1940, AdvAtPhys}. 
We also include the effect of gravity by adding the potential energy term $H_{\text{grav}} = mgz$ to the Hamiltonian of \eqnref{eq:VHamiltonian}, which typically shifts the transition by a few kHz. 
The amplitude of each \rf{} component used in the MRF APs is deduced using a co-wound pickup coil; we convert the measured voltage amplitudes into a magnetic field amplitude using the single-\rf{} Rabi frequency calibration measurements. 
The linearity of the pickup coil response was verified by repeating the single-\rf{} spectroscopy measurements for a variation in \rf{} amplitude of up to \SI{50}{\percent}.
We note that the combined MRF input approaches a value close to the saturation of the amplifier, resulting in an up to \SI{4}{\percent} compression of the amplitudes of each \rf{} component for the highest dressing \rf{} powers applied; this saturation is accounted for by the \rf{} pickup measurement.

The probe \rf{} field must be sufficiently weak that it does not itself shift the transition. For the APs used here the Rabi frequencies of the dressing \rf{}s are $100$s of \si{\kilo\Hz} while that of the probe is below \SI{100}{\Hz}. 
Selected \rf{} spectroscopy measurements were repeated with probe amplitudes spanning one third to three times its standard value, with no measurable shift of the resonance observed. 

\subsection{RF spectroscopy in the MRF potential} 
\label{sec:MRFspec}

\begin{figure*}[!t]
\includegraphics[width=0.85\textwidth]{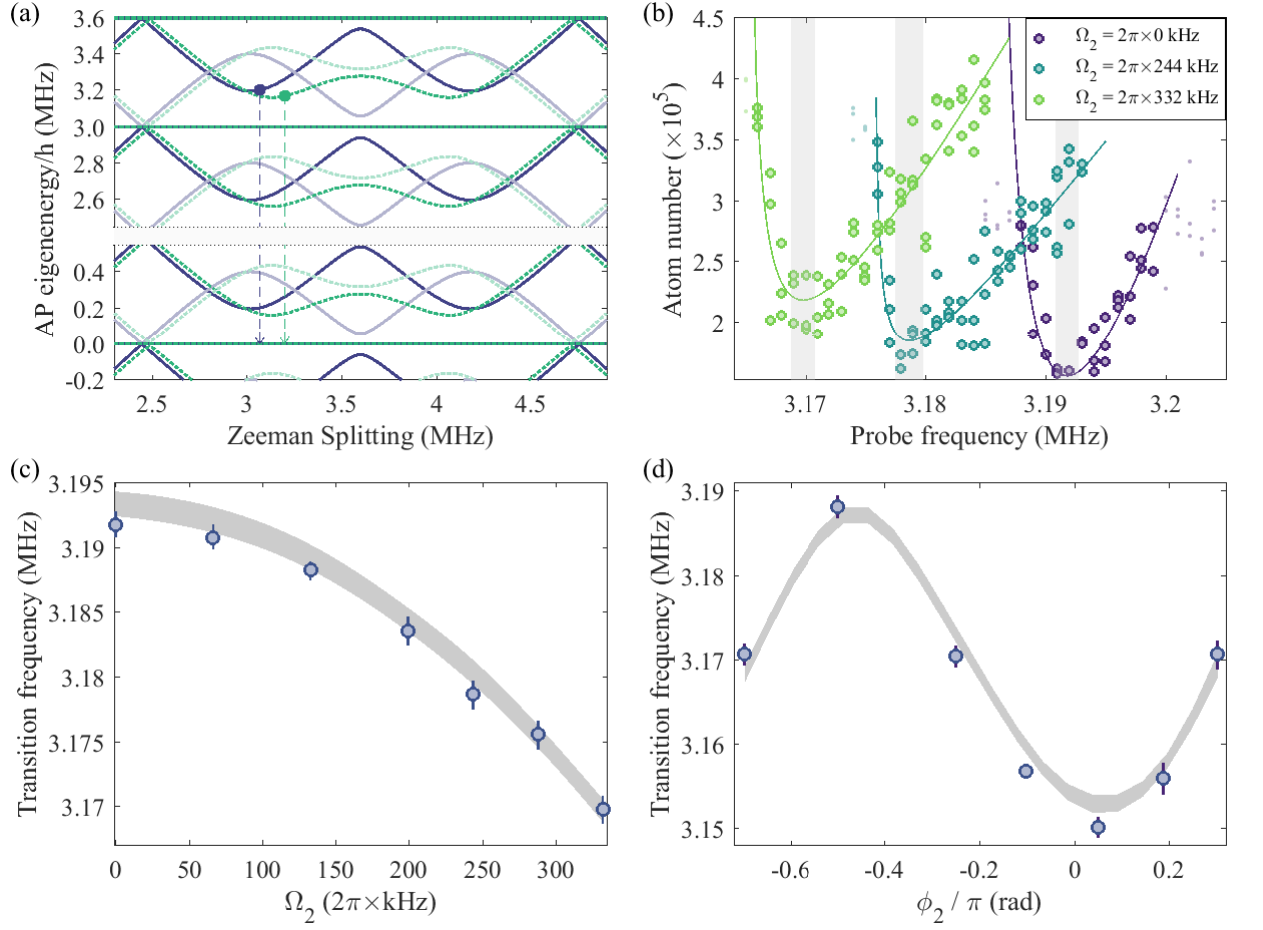}
\caption{(a) Sketch of the \rf{} spectroscopy method showing dressed state eigenenergies at two different barrier heights, plotted in the absence of gravity. The states corresponding to the two barrier heights are indicated by solid (purple) and dotted (green) lines. 
A BEC is confined in the well near \w{1}, as illustrated by the points, offset from the minima of the potential to incorporate gravity. 
We apply a probe \rf{} resonant with the dressed state transition, as illustrated by the arrows.
(b) Measured spectroscopy resonances at $\W{2} = 2\pi\times 0, 244, \SI{332}{\kilo\Hz}$, with $\W{1}=2\pi\times187$~kHz and $\W{3}=2\pi\times248$~kHz. Data points shown in bold are included in the fit used to extract the minimum of the resonance (solid lines, see text), with grey sections indicating the 99\% confidence interval for each minimum. 
(c) Change in measured (points) and theoretical (line) resonances in the MRF potential for a range of values of \W{2}, corresponding to the full data set of the resonances shown in (b). 
The phase difference between \rf{} components during this amplitude sweep was held constant, with relative phase values corresponding to the final point on plot (d) with a barrier phase term $\phi_2=0.302\pi$ radians. 
(d) Change in measured (points) and theoretical (line) resonances in the MRF potential for a range of values of $\phi_{2}$ for fixed field amplitudes $\W{1,2,3} = 2\pi\times (177, 310,245)\si{\kilo\Hz}$.
Error bars in both plots are calculated using the 99\% confidence interval in the spectroscopy resonance fit in combination with uncertainty in the \rf{} amplitude and its calibration. 
The theory line was obtained with no free parameters by calculating the transition energy for each value of \W{2} probed experimentally, with an interpolation between these values. 
Its finite width corresponds to the experimental uncertainty in the three measured \rf{} amplitudes \W{j} at each value of \W{2}. 
\label{fig:MRFspec}}
\end{figure*}

The closely spaced ladder of dressed-atom energy levels resulting from the application of multiple dressing \rf{}s leads to   
a large number of transitions between different Floquet manifolds that can be driven by an appropriate probe \rf{} field \cite{GarrawayPerrin2016, GarrawayPerrin2017}. 
However, many of these correspond to higher-order multiple-photon processes with low transition rates. 
Determining the theoretical transition frequencies begins with a calculation of the AP eigenenergies using the Floquet method of \refsec{sec:theory}, followed by selecting a single energy level corresponding to the double well from the infinite ladder of periodicity $\hbar\w{f}$. 
The condensate is localised at the position of minimum energy within the well near resonance with \w{1}.
Energy separations from this position in the trapped eigenstate to all untrapped eigenstates of the ladder are calculated, yielding a spectrum of possible transitions but with no information as to the strength of each individual transition. 

The calculated eigenenergies are experimentally verified using a BEC confined in the $\w{1}$ 
shell, using a linearly polarised MRF field to minimise experimental variables and  eliminate any experimental uncertainty arising from the phase between x and y field components. 
The spectroscopy method, calculated values, and measured results are illustrated in \reffig{fig:MRFspec}. 
We measure the dressed state transition as illustrated in \reffig{fig:MRFspec}(a). 
By separately varying \W{2} and $\phi_{2}$, the amplitude and phase of the barrier \rf{}, we experimentally probe the effects of these two parameters. 
These results are plotted in \reffig{fig:MRFspec}(c) and (d) respectively. The theoretical transitions were calculated for each set of measured \rf{} field amplitudes \W{j} and phases $\phi_{j}$, and plotted with a finite width corresponding to the uncertainty arising from quadrupole gradient and \rf{} amplitude calibrations.

The \rf{} amplitude ramps for these measurements follow a similar method to that discussed in \refsec{sec:ExpMethods} but starting with a BEC in the shell formed by the linearly polarised $\w{1}$ field component, 
ramped from circular polarisation over \SI{500}{\milli\second}. 
\W{2} and \W{3} 
are then ramped up to their final values with a set relative phase, to form the MRF potential in which \rf{} spectroscopy is performed. 
For the barrier amplitude spectroscopy measurement plotted in \reffig{fig:MRFspec}(c), $\W{1} = 2\pi\times\SI{187}{\kilo\Hz}$ and $\W{3} = 2\pi\times\SI{248}{\kilo\Hz}$, while $\rabi_2$ takes values between $0$ and $2\pi\times\SI{332}{\kilo\Hz}$ with a quadrupole gradient $B_{q}^{\prime}=\SI{82.5}{\gauss\per\centi\meter}$. 
Over the course of the \W{2} amplitude ramp, we measure a fall in \W{1} by 5\% and rise in \W{3} by 1\% due to amplifier saturation and nonlinearities. 
This amplitude sweep is performed with a fixed phase relationship between the \rf{} components, with relative phase components $\phi_{(1,2,3)}=(0,0.302\pm0.001, 0.132\pm0.002)\pi$ radians where the quoted uncertainty is given by the standard deviation of the measured relative phase of each \rf{} component. 
The measured field amplitudes and relative phase values are accounted for in the calculated transition frequencies plotted as the theoretical grey line in \reffig{fig:MRFspec}.
The phase variation measurement shown in \reffig{fig:MRFspec}(d) sees barrier amplitudes fixed at $\W{1,2,3} = 2\pi\times (177, 310,245)\si{\kilo\Hz}$ with $B_{q}^{\prime}=\SI{82.8}{\gauss\per\centi\meter}$ and $\phi_{2}$, the relative phase of the barrier component, varied over a $\pi$ range. 
The amplitudes \W{1} and \W{3} are set such that the condensate remains confined to the initial well for the spectroscopy measurements, during which the weak \rf{} probe is applied for a duration of \SI{40}{\milli\second}. 
The potential is deformed slowly to avoid sloshing of the condensate; ramps occur over an \SI{800}{\milli\second} duration that is slow compared to the inverse of the 200 to \SI{400}{\Hz} axial trap frequencies. 
The probe duration is sufficiently long that any residual sloshing in the wells would only manifest as 
a broadening of the measured \rf{} spectroscopy resonances.

The resonance point is extracted from the asymmetric spectroscopy profile~\cite{Easwaran2010} by fitting a function of the form $a(x-b)+c/\sqrt{x-d}$. 
This function provides a good approximation to the asymmetric lineshape of the  resonance profile 
from which the resonant probe frequency that minimises the atom number can be extracted. 
Only the data points lying within the range of the resonance were included in the fit, such that the asymmetric parabola captures the centre of the resonance with minimal free parameters.

The actual lineshape can be simulated numerically~\cite{Easwaran2010}, and is influenced by the amplitudes of both dressing and probe \rf{} fields, and the chemical potential of the trapped condensate. 
With these factors, a separate fit for each spectroscopy data set is impractical and at risk of overfitting. 
Qualitative comparisons between the simulated lineshape and chosen fit function suggest that the systematic uncertainty arising from a discrepancy between these models would be smaller than a kHz.
The uncertainty in the fitted resonance location for both single-\rf{} calibration and MRF potentials is estimated from the \SI{99}{\percent} confidence interval of the fitted minimum, and is of order 1 to \SI{3}{\kilo\Hz}, although with a statistical accuracy limited by the sample size. This forms the dominant source of uncertainty in the measured transition frequencies, with a smaller influence from uncertainty in measuring dressing \rf{} amplitudes with the pickup coils. 
Agreement is found with calculated values for the transition frequencies for both amplitude and phase measurements. 

The total width of each MRF spectroscopy resonance is of order 10~kHz, with the peak itself identifiable to within 3~kHz. The 40 kHz shift of the resonance peak over the full range of the parameter sweep is thus clearly resolved. The widths of each resonance are comparable to Ref.~\cite{Hofferberth2007} although broader than those presented in Ref.~\cite{Merloti2013}. This arises from the relatively weak vertical trap frequencies of 290~Hz in this work, as compared to 2~kHz in Ref.~\cite{Merloti2013}, and the consequent increase in the broadening effect of the gravitational sag.

As shown in \reffig{fig:MRFspec}, increasing \W{2} to lower the barrier reduces the energy separation between trapped and untrapped states for the measured transition. A shift in the measured \rf{} spectroscopy resonance on the order of tens of kHz is observed as $\W{2}$ is varied, in agreement with the theory. The variation in transition energy with phase $\phi_{2}$ relative to $\phi_{1,3}$, resulting from the dependence of nonlinear processes on the overall shape of the waveform, 
demonstrates a periodicity in $\pi$ expected from the numerical calculations; 
the same calculations suggest that a $2\pi$ periodicity would arise from varying $\phi_{3}$.

\section{Conclusions and outlook}
\label{sec:conclusions}
We have performed the first experimental implementation of a multiple-\rf{} adiabatic potential, using three separate dressing \rf{}s to produce a double well configuration with independent control over each trapping well and the barrier between them. 
We have demonstrated potential shaping through manipulation of the individual \rf{} amplitudes, achieving transport from one well to another, a reliable loading sequence for this double well, and dynamic control over the barrier height.
Experimental characterisation of the MRF potential by \rf{} spectroscopy of a trapped BEC validates the theoretical calculation of MRF eigenenergies by Floquet theory.

The separation of the wells in our scheme is determined by the quadrupole gradient and frequency spacing of the MRF components. 
In this work, we have demonstrated a large spacing of order \SI{100}{\micro\metre}. 

This choice was motivated by the desire to image the double well \emph{in situ} with a low NA imaging system. Far smaller separations are possible using smaller frequency intervals and higher quadrupole gradients, limited only by the constraint that atoms follow the potential adiabatically~\cite{Burrows2017}. For example, we have confined a BEC in a double well with a separation of \SI{7.5}{\micro\m}, using a frequency interval of 200 kHz, which is sufficient for matter-wave interference experiments. 
Exploiting the anisotropic character of \rf{} dressed potentials~\cite{Merloti2013}, our technique could be used to probe the behaviour of 2D systems~\cite{Mathey2010}.
Further reduction to a separation suitable for the observation of tunnelling or Josephson oscillations is possible within the constraints imposed by adiabaticity. 

Dressing with multiple independently generated radiofrequencies opens a range of new opportunities beyond the existing single-\rf{} adiabatic potential experiments while retaining their characteristic smoothness and low heating rates.
As an extension of this work, additional frequency components enable the implementation of more complex geometries such as lattices~\cite{Courteille2006}, box traps, or wells coupled to larger reservoirs.
Independent control over both the polarisation and amplitude of each \rf{} component permits further manipulations, for example to connect our two trapping potentials at different locations through the spatial variation of the coupling strength.
The MRF technique can also be combined with existing proposals to produce AP lattices using micro-structured arrays of conductors~\cite{Sinuco-Leon2015,Sinuco-Leon2016},
or provide a means of independent species-selective confinement for mixtures of atomic species with different $g_F$ values~\cite{Bentine2017}.

\section*{Acknowledgements}
The authors would like to thank Rian Hughes for comments on the manuscript. This work was supported by the EU H2020 Collaborative project QuProCS
(Grant Agreement 641277). TLH, EB, KL and AJB thank the EPSRC for doctoral training funding. 

\bibliographystyle{iopart-num}

\providecommand{\newblock}{}

\end{document}